\documentclass[prd,preprint,nofootinbib,showpacs]{revtex4-1}
\usepackage{graphicx}
\usepackage{hyperref}
\usepackage{amsfonts}
\usepackage{amsmath,amssymb}
\usepackage{bm}
\usepackage{color}

\def\be{\begin{equation}}
\def\ee{\end{equation}}
\def\bea{\begin{eqnarray}}
\def\eea{\end{eqnarray}}
\def\ba{\begin{array}}
\def\ea{\end{array}}

\begin{document}
\title{Conductivity of Strongly Coupled Striped Superconductor}

\author{Jimmy A. Hutasoit}
\email{jah77@psu.edu} 
\affiliation{Department of Physics, West Virginia University, Morgantown, West Virginia 26506}
\author{George Siopsis}
\email{siopsis@tennessee.edu}
\author{Jason Therrien}
\email{jtherrie@tennessee.edu}
\affiliation{Department of Physics and Astronomy, The University of Tennessee, Knoxville, Tennessee 37996-1200}

\date{\today} 
\pacs{11.15.Ex, 11.25.Tq, 74.20.-z}

\begin{abstract}
We study the conductivity of a strongly coupled striped superconductor using gauge/gravity duality (holography). The study is done analytically, in the large modulation regime. We show that at low temperatures, the optical conductivity is inhomogeneous and anisotropic, but with the anisotropy vanishing when the scaling dimension of the superconducting order parameter is unity. Near but below the critical temperature, we calculate the conductivity analytically at small frequency $\omega$ and find it to be both inhomogeneous and anisotropic. The anisotropy is imaginary and scales like $1/\omega$. We also calculate analytically the speed of the second sound and the thermodynamic susceptibility.
\end{abstract}

\maketitle

\section{Introduction}
Recently, there has been a flurry of activities applying gauge/gravity duality to study strongly coupled condensed matter systems (for a review, see for example, Ref. \onlinecite{Sachdev:2011fk}). A large portion of these activities revolves around understanding the so-called high-$T_c$ or high-temperature superconductors. The high-$T_c$ superconductors, such as cuprates and iron pnictides, are interesting, because not only do they exhibit strong correlation, but they also exhibit the existence of competing orders that are related to the breaking of the lattice symmetries. At first, these orders seem to be unrelated to superconductivity, however, a study of the effect of inhomogeneity of the pairing interaction in a weakly coupled BCS system \cite{Martin:2005fk} as well as numerical studies of Hubbard models \cite{Hellberg:uq,White:kx} suggest that inhomogeneity might play a role in high-$T_c$ superconductivity. Furthermore, the recent discovery of transport anomalies in ${\rm La}_{2-x} {\rm Ba}_x {\rm CuO}_4$, which are particularly prominent for $x = 1/8$ \cite{Li:2007}, might be explained under the assumption that this cuprate is a superconductor with a unidirectional charge density wave, i.e.,  a ``striped" superconductor \cite{Berg:2009fk}.  Other studies using mean-field theory have also shown that unlike the homogeneous superconductor, the striped superconductor exhibits the existence of a Fermi surface in the ordered phase \cite{Baruch:2008uq,Radzihovsky:2008kx} and its complex sensitivity to quenched disorder \cite{Berg:2009fk}.

In gauge/gravity duality, the strongly coupled condensed matter systems are mapped to a weakly coupled gravitational theory on a spacetime with negative cosmological constant (asymptotically anti de-Sitter (AdS) spacetimes). In particular, the study of strongly coupled superconductors is mapped to a study of Einstein-Maxwell-scalar theory of four-dimensional AdS black holes, and the superconducting phase transition is understood as the (unstable) black holes forming scalar 'hair.' To study the effect of inhomogeneity at strong coupling, Refs. \cite{Flauger:2011vn, Hutasoit:2012fk,Ganguli:2012fk} introduced a modulated chemical potential, which in the language of gauge/gravity duality is translated into a modulated boundary value for the electrostatic potential in the AdS Schwarzschild black hole. In particular, Ref.\ \cite{Hutasoit:2012fk} studied analytically the critical temperature and the behavior of the order parameter (condensate) as a function of the modulation wave number. 

In this article, we study the conductivity of the holographic striped superconductor by adding perturbations to the solution found in Ref.\ \cite{Hutasoit:2012fk}. Our main goal is two-fold:
\begin{itemize}
\item[$(a)$] To study in detail the anisotropic behavior of conductivity and its dependence on the temperature as well as frequency, by using analytic techniques as much as possible. As this is something that can be measured in experiment, studying the anisotropy of conductivity can lead us toward a more realistic model of high temperature superconductors using holographic techniques.
\item[$(b)$] To establish that indeed this model exhibits a superconducting transition.
\end{itemize} 
Following Ref.\ \cite{Hutasoit:2012fk}, we focus on the large modulation regime for an analytical study. A more thorough analysis requires a combination of analytic and numerical techniques and will be taken up elsewhere. Our study is conducted at temperatures near the critical temperature $T \approx T_c$, and at low temperature $T\ll T_c$. In both regimes, we find that the dynamical conductivity does not only depend on the frequency and wavelength of the electric field applied to the strongly coupled system, but also has a spatial dependence, as expected. More importantly, we find that at $T \leq T_c$, at zero wavelength, the real part of the conductivity contains a term that is proportional to $\delta(\omega)$, where $\omega$ is the frequency of the applied electric field. Since our system is not translationally invariant by construction, this indeed shows that the system exhibits superconductivity below the critical temperature. Furthermore, at $T \lesssim T_c$, we find that the optical conductivity is generally anisotropic. At small frequency, the anisotropy is imaginary and scales like $1/\omega$. At higher frequencies, a numerical calculation is required. We shall report on the intermediate and high frequency regimes in the near future. At low temperature, and scaling dimension of the superconducting order parameter $\Delta =1$, we were able to derive explicit analytic expressions for the conductivity, and showed that the conductivity is isotropic. Even though the general analytic expressions for $\Delta \ne 1$ are too cumbersome to write explicitly, we were able to show that the anisotropy is non-vanishing in this case. A more extensive study using numerical techniques in this regime is in progress. Lastly, at $T\approx T_c$, we also calculated the speed of the second sound and the thermodynamic susceptibility.

Our discussion is organized as follows. In Section \ref{sec:2}, we set up the system and fix our notation. In Section \ref{sec:3}, we obtain analytic results for the conductivity at low temperatures. In Section \ref{sec:4} ,we calculate the conductivity analytically near the critical temperature. It should be noted that the analytical techniques used in the two temperature regimes discussed in Sections \ref{sec:3} and \ref{sec:4}, respectively, differ from each other. Finally, Section \ref{sec:5} contains our concluding remarks.

\section{Set-up}
\label{sec:2}
For completeness, let us briefly review the set-up and the results of  Ref.\ \cite{Hutasoit:2012fk}. The minimal requirement to study strongly coupled superconductor using gauge/gravity duality is to have a scalar field and a $U(1)$ gauge field living in a spacetime with negative cosmological constant. The scalar field is dual to a superconducting order parameter, i.e., the condensate, while the $U(1)$ gauge field is dual to the current in the condensed matter system. 

To study the strong coupling regime of the superconductor, we only need to study the gravity theory at the classical level. In particular, we are interested in finding solutions to the classical equations of motion whose boundary values are related to the parameters of the superconductor. In the large modulation regime, the backreaction can be neglected \cite{Ganguli:2012fk} and doing so, the problem of studying a strongly coupled superconductor with an effective two spatial dimensions is reduced to studying the dynamics of a scalar field with mass $m$, coupled to a $U(1)$ gauge field in the background of a 3+1-dimensional AdS Schwarzschild black hole with the following metric
\be\label{eq:1}
ds^2 = \frac{1}{z^2} \left(-  h(z)\, dt^2 + \frac{dz^2}{h(z)} + d\vec{x}^2\right) \ \ , \ \ \ \
h = 1 - z^3~,
\ee
in units in which the AdS radius is unity. 
In this coordinate system, the boundary is at $z=0$, while the horizon is at $z=1$.

The equation of motion for the scalar field is
\be\label{eq:eompsi}
- \frac{1}{\sqrt{-g}}D_a\left(\sqrt{-g}g^{ab}D_b\psi\right) + \frac{1}{2} \frac{\psi}{|\psi|} V'(|\psi|) = 0 \,,
\ee
while the equation for the U(1) gauge field is
\be\label{eq:eomA}
\frac{1}{\sqrt{-g}}\partial_a\left(\sqrt{-g} F^{ab}\right) = i g^{ab} \left[\psi^* D_a\psi - \psi (D_a\psi)^* \right]
\,,
\ee
where $a,b=t,x,y$ are space-time indices, $D_a= \partial_a - i A_a$, and $F=dA$ is the field strength of the $U(1)$ potential $A_a$. The background metric $g_{ab}$ is given by \eqref{eq:1}. We shall consider a scalar field of mass $m$, setting $V(|\psi|) = m^2 |\psi|^2$.

We introduce inhomogeneity by including a modulated chemical potential of the form
\be \mu (x) = \mu (1-\delta) + \mu \delta \cos Qx~. \ee
Thus, the charge density wave in the system is being sourced by a modulated chemical potential. The electrostatic potential is then given by 
\be
A_t(z,x) = \mu \sum_n \mathcal{A}^{(n)} (z) \cos nQx~,
\ee 
obeying the boundary condition $A_t (0,x) = \mu (x)$, which implies $\mathcal{A}^{(0)} (0) = 1-\delta$, $\mathcal{A}^{(1)} (0) = \delta$, and $\mathcal{A}^{(n)} (0) =0$ for $n\ge 2$. We are particularly interested in the regime where the modulation wave number $Q$ is a lot larger than the other scales in the system, {\it i.e.}, $T_c$ and $\mu$. In this regime, the mode $\mathcal{A}^{(1)}$ is exponentially suppressed compared to $\mathcal{A}^{(0)}$ in the bulk. The higher modes $\mathcal{A}^{(n)}$, with $n > 1$, vanish identically at $T \geq T_c$ and obtain a small but non-zero value $\mathcal{A}^{(n)} \ll \mathcal{A}^{(1)}$ at $T<T_c$.

The solution for the scalar field was analyzed in detail in Ref.\ \cite{Hutasoit:2012fk}, where a second-order transition was found at a critical temperature $T_c$ below which a condensate formed. Here we shall use the results obtained in \cite{Hutasoit:2012fk} to analyze the behavior of conductivity.

At $T \geq T_c$, the order parameter, and thus the scalar field, vanishes, but below $T_c$, we have a non-vanishing scalar field. Separating real and imaginary parts, $\psi = \Psi + i\chi$, we adopt a gauge in which $\chi =0$, and
\be
\Psi = \frac{\langle {\cal O}_{\Delta} \rangle^{(0)}}{\sqrt{2}} \, z^{\Delta} \, \sum_{n  \geq 0} F^{(n)}(z) \, \cos nQx, \label{defmode}
\ee
with $F^{(0)}(z=0)=1$. The order parameter is
\be
\langle {\cal O}_{\Delta} \rangle = \sum _{n  \geq 0} \,  \langle {\cal O}_{\Delta} \rangle^{(n)} \, \cos nQx~,
\ee
with 
\be
\langle {\cal O}_{\Delta} \rangle^{(n)} = \langle {\cal O}_{\Delta} \rangle^{(0)} \, F^{(n)}(z=0).
\ee 

Following Ref. \cite{Horowitz:2008uq}, we define the ``energy gap" derived from the order parameter by
\be \mathcal{E}_g = [\langle {\cal O}_{\Delta} \rangle]^{1/\Delta}~. \ee
We note that this quantity has the dimension of energy and in the free theory limit reproduces the gap of free/weakly coupled Landau-Ginzburg theory of superconductivity. In the homogeneous free theory, $\mathcal{E}_g$ is equivalent to the ``hard" gap of the BCS theory below which there are no quasiparticle excitations. Unfortunately, in the holographic approach discussed here, a microscopic description (similar to the one in the BCS theory) is absent, and so it is not clear what the exact relationship between $\mathcal{E}_g$ and the quasiparticle spectrum is in general. It is known that in the homogeneous free theory, $\mathcal{E}_g$ is half of the gap derived from the real part of the optical conductivity, as expected if the gapped charged quasiparticles are produced in pairs. However, in holographic homogeneous superconductors, this ratio deviates from $1/2$ \cite{Horowitz:2008uq,Siopsis:2010uq}.

The scaling dimension of the superconducting order parameter ${\cal O}_{\Delta}$ is given by
\be \Delta = \Delta_\pm = \frac{3}{2} \pm \sqrt{\frac{9}{4} + m^2} \ee
with $m$ being the mass of the scalar field dual to the order parameter. We shall examine the range
\be\label{eq8}
\frac12 < \Delta \le \frac32,
\ee
{\em i.e.,} $\Delta = \Delta_-$, corresponding to masses in the range $0 > m^2 \ge -\frac{9}{4}$.
At large $Q$, the higher modes of the scalar order parameter are suppressed compared to the zero mode,
\be
\frac{\langle {\cal O}_{\Delta}\rangle^{(n)}}{\langle {\cal O}_{\Delta}\rangle^{(0)}} \leq {\cal O}\left(\frac{\mu^{2 \left[n/2\right]}}{Q^{2 \left[n/2\right]}}\right),
\ee
where $[n]$ denotes the smallest integer $\geq n$. Furthermore, when the homogeneous part of the chemical potential vanishes , {\it i.e.}, $\delta = 1$, the odd modes vanish. 

We can now consider adding the vector potential as a perturbation on top of this background. From the on-shell Maxwell action, we can then read the current-current retarded correlation which will then give us the conductivity.

\section{Low Temperature}
\label{sec:3}

In this section, we study the behavior of the conductivity analytically at low temperature. It should be noted that in the range of scaling dimensions of the order parameter \eqref{eq8}, the condensate $\langle\mathcal{O}_\Delta \rangle$ diverges in the zero temperature limit, signaling the breakdown of the probe limit approximation we are working in. Therefore, one must be careful in the definition of the low temperature regime. The probe  limit is obtained as the charge of the scalar field diverges ($q\to\infty$). The two limits, $T\to 0$ and $q\to\infty$, are not interchangeable. The charge was set to $q=1$ in Ref.\ \cite{Hutasoit:2012fk}. Restoring it, we find that we may use the probe limit approximation for $\langle\mathcal{O}_\Delta \rangle \ll q$ \cite{Siopsis:2010fk}. Thus, the low temperature regime in which it is safe to use the probe limit approximation is given by 
\be 1 \lesssim \langle\mathcal{O}_\Delta \rangle \ll q~.\ee

\subsection{Conductivity parallel to the direction of the stripes}
First, let us study the conductivity $\sigma_y$, which is parallel to the direction of the stripes, at low temperature. To do so, we introduce a small perturbation 
\be\label{eq10}
A_y  = \int \frac{d\omega \, d^2 k}{(2\pi)^3} \, a_y(z; \omega, k) \, e^{i (kx - \omega t)}.
\ee
The leading order of the equation of motion \eqref{eq:eomA} is then given by
\be \partial_z \left(h \partial_z a_y\right) + \frac{\omega^2}{h} a_y = \left(k^2 + {\langle {\cal O}_{\Delta} \rangle^{(0)}}^2 \, z^{2(\Delta-1)} {F^{(0)}}^2\right)  a_y, \label{eom} \ee
with an incoming\footnote{``Incoming" here means that any perturbation introduced in the spacetime will fall into the horizon of the black hole. This boundary condition corresponds to the retarded correlators of the boundary theory.} boundary condition at the horizon. We can rewrite this in Schr\"odinger form as
\be
- \partial_{z_{\ast}}^2 a_y + V  a_y = \omega^2  a_y,
\ee
where the ``potential" is given by
\be
V = V_0^{(0)} + k^2\, V_1, \ee
with
\be  V_0^{(0)} = {\langle {\cal O}_{\Delta} \rangle^{(0)}}^2 \, z^{2(\Delta-1)}h{F^{(0)}}^2\ , \ \ \
 V_1 =  h~, \ee
and we have introduced the tortoise coordinate
\be
z_{\ast} =  \int_0^z \frac{dz}{h} = \frac{1}{6} \log \frac{(1-z^3)}{(1-z)^3} + \frac{1}{\sqrt{3}} \left(\tan^{-1} \frac{1+2z}{\sqrt{3}} -\frac{\pi}{6}\right)~.
\ee
Near the boundary, $z_{\ast} \approx z \rightarrow 0$, and the horizon is at $z_{\ast} \rightarrow \infty$.

To account for incoming boundary condition at the horizon, we write
\be a_y (z; \omega ,k) = e^{i\omega z_\ast}\, \hat{a}_y (z ; \omega , k), \ee
where $\hat{a}_y$ is regular at the horizon. The equation of motion for $\hat{a}_y$ is then given by
\be - h\partial_{z}^2 \hat{a}_y +(2i\omega -h')\partial_{z} \hat{a}_y+ \frac{V}{h} \hat{a}_y = 0, \ee
and its regularity  at the horizon is ensured by the boundary condition
\be\label{eqbch} \left. (3+2i\omega) \partial_z\hat{a}_y + \frac{V}{h} \hat{a}_y \right|_{z=1} = 0. \ee
We note that $V/h$ is finite in the limit $z\to 1$.

At low temperatures, this equation can be further simplified by scaling $z \to z /\langle\mathcal{O}_\Delta \rangle^{1/\Delta}$, which places the horizon at $z= \langle\mathcal{O}_\Delta \rangle^{1/\Delta}$, and then formally expanding in the small parameter $1/\langle\mathcal{O}_\Delta \rangle^{1/\Delta}$.
At lowest order, we obtain
\be\label{eq19} - \partial_{z}^2 \hat{a}_y - 2i\hat\omega \partial_{z} \hat{a}_y + \hat V \hat{a}_y =0 \ \ , \ \ \ \
\hat V = z^{2(\Delta -1)} + \hat{k}^2 + \dots ~, \ee
where we have similarly scaled the input parameters $\omega$ and $k$, respectively,
\be \hat\omega = \frac{\omega}{\langle\mathcal{O}_\Delta \rangle^{1/\Delta}} \ \ , \ \ \ \
\hat k = \frac{k}{\langle\mathcal{O}_\Delta \rangle^{1/\Delta}}. \ee
In general, there are two linearly independent solutions distinguished by their asymptotic behavior at large $z$,
\be \hat{a}_y^\pm \sim e^{\pm z^\Delta /\Delta}, \ee
and the general solution is going to be given by a linear combination of the two
\be \hat{a}_y = c_+ \hat{a}_y^+ + c_- \hat{a}_y^-. \ee
Imposing the boundary condition \eqref{eqbch}, we deduce that the ratio of the coefficients is exponentially small,
\be \frac{c_+}{c_-} \sim e^{-2\langle\mathcal{O}_\Delta\rangle /\Delta} \ll 1 ~,\ee
and therefore, the contribution of $\hat{a}_y^+$ is negligible at low temperatures.

The equation of motion can be solved analytically in the high frequency limit as a series expansion, by treating the potential term as perturbation. We obtain
\be\label{eq23} \hat{a}_y = 1 - \frac{i\hat{k}^2 }{2\hat{\omega}} z + \frac{i}{2\hat\omega (2\Delta -1)} z^{2\Delta -1} + \frac{1}{(2i\hat\omega)^{2\Delta}} \left[ \Gamma (2\Delta -1) - e^{-2i\hat\omega z} \Gamma (2\Delta -1,- 2i\hat\omega z) \right] + \dots \ee
from which we deduce the low-temperature conductivity in the high-frequency limit
\be\label{eq24} \sigma_y (\omega ,k) = 1 - \frac{\hat{k}^2}{2\hat\omega^2} +\frac{2\Gamma (2\Delta -1)}{(-2i\hat\omega)^{2\Delta}} +\dots \ee
To see the behavior at low frequencies, let us set $k=0$ and calculate the optical conductivity. This time, the term proportional to the frequency can be treated as perturbation. We obtain the solution at lowest order in $\hat\omega$,
\be \hat{a}_y = \sqrt{z} K_{\frac{1}{2\Delta}} \left( \frac{z^\Delta}{\Delta} \right) + \mathcal{O} (\hat\omega), \ee
from which we deduce the low frequency optical conductivity
\be\label{eq26} \sigma_y (\omega , k=0) \approx 
\frac{\Gamma(- \frac{1}{2\Delta})}{ (2\Delta)^{1/\Delta} \Gamma (\frac{1}{2\Delta})} \langle \mathcal{O}_\Delta \rangle^{1/\Delta}\, \frac{i}{\omega}~. \ee
%
Using the Kramers-Kronig relations, we can then deduce the real part of the DC conductivity,
\be\label{eq38a}
\Re  \left[\sigma_y(\omega = 0, k=0, x)\right]  = \frac{\pi\Gamma(- \frac{1}{2\Delta})}{ (2\Delta)^{1/\Delta} \Gamma (\frac{1}{2\Delta})} \langle \mathcal{O}_\Delta \rangle^{1/\Delta} \delta(\omega)~.
\ee
It should be noted that this result by itself does not show that we are in a superconducting phase. An infinite DC conductivity can be the result of translational invariance in the $y$-direction (along the stripes). Superconductivity will be established by deriving a similar result in the $x$-direction (perpendicular to the stripes), along which translational invariance is broken.

We can also solve the equation of motion analytically for $\Delta =1$ and $\Delta = 3/2$. For $\Delta = 1$, $\hat V$ is constant. We obtain the solution
\be \hat{a}_y \approx \hat{a}_y^- = e^{\left( -i\hat\omega - \sqrt{\hat V - \hat\omega^2 } \right) z}, \ee
from which, by using the Kubo formula, we obtain the low-temperature conductivity for $\Delta =1$,
\be\label{eq28}
\sigma_y (\omega, k) =\sqrt{1 - \frac{\hat V}{\hat\omega^2}} = \sqrt{1 - \frac{\langle\mathcal{O}_\Delta\rangle^2 + k^2}{\omega^2} }~,
\ee
in agreement with our earlier result \eqref{eq24} for $\Delta =1$ in the high frequency limit. The expression \eqref{eq28} also agrees with the low-frequency optical conductivity \eqref{eq26} for $\Delta =1$.

For $\Delta = \frac{3}{2}$, the solution is given in terms of the Airy function,
\be \hat{a}_y \approx \hat{a}_y^- = e^{-i\hat\omega z} \mathrm{Ai} (\hat{k}^2 - \hat{\omega}^2 + z) .\ee
We deduce the low-temperature conductivity for $\Delta = \frac{3}{2}$,
\be\label{eq30} \sigma_y (\omega , k) = \frac{i\mathrm{Ai}' (\hat{k}^2 - \hat{\omega}^2)}{\hat\omega\mathrm{Ai} (\hat{k}^2 - \hat{\omega}^2)}, \ee
in agreement with the low-frequency optical conductivity \eqref{eq26} for $\Delta = \frac{3}{2}$. As we increase $\omega$, this expression develops poles and cannot be trusted.

For general $\Delta$, we can solve the equation of motion 
by approximating the potential using the procedure introduced in Ref.\ \cite{Siopsis:2010uq}. We replace the potential $\hat{V}$ by a constant $\mathcal{V}_0$, thus approximating the wave equation \eqref{eq19} by
\be\label{eq19a} -\partial_z^2 \hat{a}_y -2i\hat{\omega} \partial_z\hat{a}_y + \mathcal{V}_0 \hat{a}_y = 0~. \ee
The constant $\mathcal{V}_0$ is determined in a self-consistent manner as follows.
First we obtain the solution to the approximate wave equation \eqref{eq19a},
\be
\hat{a}_y \approx  \hat{a}_y^- \approx e^{\left( -i\hat\omega - \sqrt{\mathcal{V}_0 - \hat\omega^2 } \right)z}~, \label{incomingsol1}
\ee
and then replace the constant $\mathcal{V}_0$ by the average potential $\langle \hat{V} \rangle$ which is obtained using the approximate solution \eqref{incomingsol1}.
The approximate potential $\mathcal{V}_0 = \langle \hat{V}\rangle$ is frequency-dependent and provides a good approximation for all $\omega$ \cite{Siopsis:2010uq}.

In particular, at zeroth order, we obtain the self-consistency condition which determines $\langle \hat{V} \rangle$,
\be\label{potential0}
\langle \hat{V}^{(0)} \rangle = \frac{\int_0^\infty dz_{\ast} \hat{V}^{(0)} \left|a_y\right|^2}{\int_0^\infty dz_{\ast} \left|a_y\right|^2}
\approx \hat{k}^2 + \Gamma(2\Delta -1) \left(2 \sqrt{\langle \hat{V}^{(0)} \rangle - \hat{\omega}^2 } \right)^{2(1-\Delta)}, 
\ee
where the integrals are evaluated by giving $\omega$ an imaginary part, which is set to zero at the end.

We then obtain the low-temperature conductivity,
\bea
\sigma_y (\omega, k) &=&\sqrt{1 - \frac{\langle \hat{V} \rangle}{\hat\omega^2}},
\eea
from which we deduce
\be\label{potential} \hat\omega^2 \sigma_y^2 + \Gamma (2\Delta -1) (2i\hat\omega \sigma_y)^{2(1-\Delta)} = \hat\omega^2 - \hat{k}^2. \ee
This reproduces our earlier result \eqref{eq28} for $\Delta = 1$ exactly.

For $\Delta = \frac{3}{2}$, we obtain
\be \sigma_y = \frac{1}{i\hat\omega} \left[ \left( \frac{1}{4} + \sqrt{\frac{1}{16} + \left( \frac{\hat\omega^2 -\hat{k}^2}{3} \right)^3} \right)^{1/3} - \left( -\frac{1}{4} + \sqrt{\frac{1}{16} + \left( \frac{\hat\omega^2 -\hat{k}^2}{3} \right)^3} \right)^{1/3} \right]. \ee
This is in excellent agreement numerically with our earlier result \eqref{eq30} for $\omega^2 < k^2$, even though the two expressions differ analytically. The two results start to diverge significantly for $\hat\omega^2 - \hat k^2 \gtrsim 1$.

At high frequencies, we obtain from Eq. (\ref{potential}) the conductivity
\be \sigma_y = 1 - \frac{\hat{k}^2}{2\hat\omega^2} + \frac{2\Gamma (2\Delta -1)}{(-2i\hat\omega)^{2\Delta}} + \dots \ee
in excellent agreement with our earlier result (\ref{eq24}) obtained by perturbation theory.

We note that the approximation we used to evaluate the integrals in Eq. (\ref{potential0}) renders this approach insensitive to the details of the horizon in the bulk, which corresponds to the infrared regime in the boundary theory. Therefore, this approach is valid when the gap is a lot larger than the temperature, $\mathcal{E}_g = [\langle {\cal O}_{\Delta} \rangle^{(0)}]^{1/\Delta} \gg 1$ (in our notation, the gap is measured in units of temperature, see Section \ref{sec:2}).

Now, let us include the first order correction into the equation of motion (\ref{eom}). For $\delta \ne 1$, we have
\bea\label{eq37}
\mathcal{L} \hat{a}_y(z; \omega,k) &=& \left[ k^2 + {\langle {\cal O}_{\Delta} \rangle^{(0)}}^2 \, z^{2(\Delta-1)} {F^{(0)}}^2\right]  \hat{a}_y(z; \omega,k) \nonumber \\
& & + {\langle {\cal O}_{\Delta} \rangle^{(0)}}^2 z^{2(\Delta-1)} F^{(0)} F^{(1)}  \left[ \hat{a}_y(z; \omega,k-Q)+ \hat{a}_y(z; \omega,k+Q) \right]  \nonumber \\
& & + {\langle {\cal O}_{\Delta} \rangle^{(0)}}^2 z^{2(\Delta-1)} F^{(0)} F^{(2)}  \left[ \hat{a}_y(z; \omega,k-2Q)+ \hat{a}_y(z; \omega,k+2Q) \right]~,
\eea
where
\be \mathcal{L} \hat{a}_y \equiv -\partial_z \left( h \partial_z \hat{a}_y \right) + 2i\omega \partial_z \hat{a}_y. \ee
This equation can be solved by treating the right-hand side as perturbation, as before.
In the high frequency limit, we obtain ({\em cf.}\ Eq.\ \eqref{eq23})
\bea \hat{a}_y (z; \omega,k) &=& \left[ 1 - \frac{ik^2 }{2\omega} z + {\langle \mathcal{O}_\Delta \rangle^{(0)}}^2 \mathcal{F} (z; \omega ) \right] a_y(z=0; \omega , k)\nonumber\\
&& + \langle \mathcal{O}_\Delta \rangle^{(0)}\langle \mathcal{O}_\Delta \rangle^{(1)} \mathcal{F} (z;\sqrt{\omega^2 - (k-Q)^2}) a_y (z=0;\omega , k-Q) \nonumber\\
&& + \langle \mathcal{O}_\Delta \rangle^{(0)}\langle \mathcal{O}_\Delta \rangle^{(2)} \mathcal{F} (z;\sqrt{\omega^2 -(k-2Q)^2}) a_y (z=0;\omega , k-2Q) \nonumber\\
& & + (Q\to -Q), \eea
where
\be \mathcal{F} (z; \omega ) \equiv \frac{i}{2\omega (2\Delta -1)} z^{2\Delta -1} + \frac{1}{(-2i\omega)^{2\Delta}} \left[ \Gamma (2\Delta -1) - e^{-2i\omega z} \Gamma (2\Delta -1,- 2i\omega z) \right]. \ee
The on-shell action then becomes
\be I = \int \frac{d\omega \, dk\, dk'}{(2\pi)^3} \, a_y(z=0;\omega,k) i\mathcal{G} (\omega; k,k') a_y (z=0; -\omega, k') ~, \ee
where
\be \mathcal{G} (\omega; k,k')  = g^{(0)} \delta(k'+k)
 + \langle {\cal O}_{\Delta} \rangle^{(0)}  \langle {\cal O}_{\Delta} \rangle^{(1)} g^{(1)}(\omega; k,k')
+ \langle {\cal O}_{\Delta} \rangle^{(0)}  \langle {\cal O}_{\Delta} \rangle^{(2)} g^{(2)}(\omega; k,k') ~,
\ee
and
\bea g^{(0)} &=& 1- \frac{k^2}{2\omega^2} + \frac{2\Gamma (2\Delta -1)}{(-4\omega^2)^{\Delta}} {\langle \mathcal{O}_\Delta \rangle^{(0)}}^2 \nonumber\\
g^{(1)} &=& \frac{2\Gamma (2\Delta -1)}{(-4[\omega^2 - (k-Q)^2])^{\Delta}} \delta(k'+k-Q) 
 + (Q\to -Q) \nonumber\\
g^{(2)} &=& \frac{2\Gamma (2\Delta -1)}{(-4[\omega^2 - (k-2Q)^2])^{\Delta}} \delta(k'+k-2Q) 
 + (Q\to -Q) ~,
\eea
from which we deduce the retarded current-current correlation function
\be
\langle J(-\omega,x) \, J(\omega,k) \rangle =  \frac{i}{2\pi} \int \frac{dk'}{2\pi} e^{-ik'x} \mathcal{G} (\omega ; k,k') ~. \ee The dynamical conductivity is then given by the ``pre-averaging" Kubo formula (see, e.g., \cite{Mahan})
\be
\sigma_y (\omega, k, x) = (2\pi)^2 \frac{e^{-ikx}}{i \omega} \, \langle J(-\omega,x) \, J(\omega,k)\rangle ~. \ee
Here, we use the ``pre-averaging" Kubo formula because there is interesting \emph{macroscopic} spatial dependence in the conductivity. Unlike atomic fluctuations, we do not want to get rid of this spatial dependence by averaging over $x$. For $\delta = 1$, the above expression simplifies further due to the fact $\langle {\cal O}_{\Delta} \rangle^{(1)}$ vanishes. Lastly, from the dynamical conductivity, we can obtain the optical conductivity by simply setting $k=0$.

We deduce the optical conductivity
\bea
\sigma_y (\omega, k=0, x) &\approx& 1 + \frac{\Gamma (2\Delta -1)}{2^{2(\Delta-1)}}\langle {\cal O}_{\Delta} \rangle^{(0)}
 \left[ \frac{\langle {\cal O}_{\Delta} \rangle^{(0)}}{(-\omega^{2})^{\Delta}}\right. \nonumber\\
&& \left. + 2\frac{\langle {\cal O}_{\Delta} \rangle^{(1)}}{(Q^2-\omega^2)^\Delta} \cos Q x
+ 2\frac{\langle {\cal O}_{\Delta} \rangle^{(2)}}{ (4Q^2 -\omega^2)^\Delta} \cos 2Q x \right] ~ ,
\label{condyhigh}
\eea
for $\delta \ne 1$. Again, this simplifies further for $\delta = 1$ as the odd modes of the order parameter vanish.

A more general expression, valid at both high and low frequencies, can be found in the case $\Delta =1$. We obtain
\be
\sigma_y (\omega, k, x) = \frac{1}{\omega} \left[ \Omega_k + \frac{\langle {\cal O}_{1} \rangle^{(0)}  \langle {\cal O}_{1} \rangle^{(1)}}{Q} \sigma_y^{(1)}(\omega, k,x)
+ \frac{\langle {\cal O}_{1} \rangle^{(0)}  \langle {\cal O}_{1} \rangle^{(2)}}{4Q} \sigma_y^{(2)}(\omega, k,x) \right] ~,
\ee
\bea
\sigma_y^{(1)}(\omega, k,x) &=&  \frac{ \Omega_{ k-Q} - \Omega_k}{Q - 2 k} e^{-iQx} 
 + \frac{\Omega_{ k+Q} - \Omega_k}{Q + 2 k} e^{iQx}, \nonumber \\
\sigma_y^{(2)}(\omega, k,x) &=& \frac{ \Omega_{ k-2Q} - \Omega_k}{Q -  k} e^{-2iQx}
 + \frac{\Omega_{ k+2Q} - \Omega_k}{Q +  k} e^{2iQx} ~. \eea

Since $F^{(1)}$ and $F^{(2)}$ have $Q$-dependence suppression, we can solve the wave equation \eqref{eq37} perturbatively,
\be
\hat{a}_y(z;\omega,k) = \hat{a}^0_y(z;\omega,k) + \hat{a}^1_y(z;\omega,k),
\ee
with boundary condition
\be
\hat{a}_y (z=0 ; \omega ,k) = a_y(z=0;\omega,k) = \hat{a}^0_y(z=0;\omega,k)~.
\ee
Working as before, for $\Delta =1$ we obtain at zeroth order
\be 
\hat{a}^0_y(z;\omega,k) = \hat{a}^0_y(z=0;\omega,k) \, e^{-i[\omega- \Omega_k] z} \ , \ \ \ \Omega_k \equiv \sqrt{\omega^2 - {\langle {\cal O}_{1} \rangle^{(0)}}^2 - k^2} \, ,
\ee
and at first order,
\bea
\hat{a}^1_y(z;\omega,k) &=& \frac{ {\langle {\cal O}_{1} \rangle^{(0)}}^2}{Q} \langle V_0^{(1)} \rangle \frac{\hat{a}^0_y(z=0;\omega, k+Q)}{Q + 2 k} \left(e^{i \Omega_{k+Q}z}- e^{i \Omega_k z}\right) \nonumber \\
& & + \frac{{\langle {\cal O}_{1} \rangle^{(0)}}^2}{4Q} \langle V_0^{(2)} \rangle \frac{\hat{a}^0_y(z=0;\omega, k+2Q)}{Q +  k} \left(e^{i \Omega_{ k+2Q} z}- e^{i \Omega_k z}\right) \nonumber\\
& & + (Q\to -Q)~, \label{eq:momshift}
\eea
where 
\be
\langle V_0^{(i)} \rangle = \frac{\int_0^\infty dz_{\ast}\,F^{(i)}h\left|a_y\right|^2}{\int_0^\infty dz_{\ast}\, \left|a_y\right|^2}
\approx \frac{\langle {\cal O}_{1} \rangle^{(i)}}{\langle {\cal O}_{1} \rangle^{(0)}} \langle V_0^{(0)} \rangle \ \  \ \ \ \  (i=1,2)~,
\ee
and we have approximated the integral by evaluating the integrand near the boundary. 

For $\Delta \ne 1$, we can solve Eq.\ \eqref{potential} at the high-frequency regime $\omega \gtrsim Q, [\langle {\cal O}_{1} \rangle^{(0)}]^{1/\Delta}$, giving the optical conductivity
\be
\sigma_y (\omega, k=0, x) \approx 1 + \frac{(-i\omega_0)^{2\Delta}}{2 \omega^{2\Delta}}
 \left[ 1+ 2\frac{\langle {\cal O}_{\Delta} \rangle^{(1)}}{ \langle {\cal O}_{\Delta} \rangle^{(0)}} \cos Q x
+ 2\frac{\langle {\cal O}_{\Delta} \rangle^{(2)}}{ \langle {\cal O}_{\Delta} \rangle^{(0)}} \cos 2Q x \right] ~ ,
\label{condyhigh1}
\ee
where
\be\label{eqo0} \omega_0 = \left(2^{2(1-\Delta)}\,\Gamma(2\Delta-1) \, {\langle {\cal O}_{\Delta} \rangle^{(0)}}^2\right)^{\frac{1}{2\Delta}} ~, \ee
for $\delta \ne 1$. Again, this simplifies further for $\delta = 1$ as the odd modes of the order parameter vanish.

We can also solve Eq.\ \eqref{potential} in the low frequency regime. We obtain an expression similar to \eqref{eq26},
\be \sigma_y (\omega, k=0, x) \approx \left[ 1+ \frac{\langle {\cal O}_{\Delta} \rangle^{(1)}}{ \Delta\langle {\cal O}_{\Delta} \rangle^{(0)}} \cos Q x
+ \frac{\langle {\cal O}_{\Delta} \rangle^{(2)}}{ \Delta\langle {\cal O}_{\Delta} \rangle^{(0)}} \cos 2Q x \right] \frac{i\omega_0}{\omega}~,
\label{condy}
\ee
for $\delta \ne 1$. As before, for $\delta = 1$, the odd modes of the order parameter vanish. We note that this leading term in the small frequency expansion of the optical conductivity vanishes when the scalar order parameter vanishes (i.e., $\omega_0 = 0$, from Eq.\ \eqref{eqo0}), which corresponds to the onset of superconductivity, as we will show.

\subsection{Conductivity perpendicular the direction of the stripes}
As noted in Ref. \cite{Flauger:2011vn}, because the inhomogeneity is in the $x$-direction, applying an electric field in the $x$-direction sources other independent perturbations even at linear order, making the computation of $\sigma_x$ complicated. The perturbations that need to be turned on are
\bea
A_x  &=& a_x(z, x) e^{i\omega z_\ast} e^{i (kx - \omega t)}, \nonumber \\
A_t  &\rightarrow& A_t +  a_t(z, x) e^{i\omega z_\ast}e^{i (kx - \omega t)}, \nonumber \\
\Psi &\rightarrow& \Psi \left[ 1 +  f_1(z, x) e^{i\omega z_\ast} e^{i (kx - \omega t)} \right]~, \nonumber \\
\chi &=&  \Psi  f_2(z, x) e^{i\omega z_\ast} e^{i (kx - \omega t)}.
\eea
Here, adding the small perturbation in term of the bulk vector potential sources perturbations for the bulk electrostatic potential and for both the real and imaginary part of the bulk scalar field, $\Psi$ and $\chi$. We have also let the Fourier coefficients $a_x$, $a_t$, $f_1$ and $f_2$ retain residual $x$-dependence due to the fact that the subleading solutions are related to the leading order solutions whose momenta differ by multiples of $Q$. This is exactly what we saw in the previous subsection (see Eq.\ (\ref{eq37})) which ultimately leads to the spatial dependence of the conductivity. Alternatively, one can express the perturbation as a Fourier transform, as we did in the case of transverse perturbations (Eq.\ \eqref{eq10}). We also included a convenient factor $e^{i\omega z_\ast}$ to account for the incoming wave behavior at the horizon.

The equations of motion are then given by
\bea\label{eq60}
\partial_z^2 f_1 +\left[ \frac{2i\omega +h'}{h} + \frac{2\partial_z\Psi}{\Psi} - \frac{2}{z} \right] \partial_z f_1 + \frac{2}{h} \frac{\partial_x\Psi}{\Psi} \partial_x f_1+ \frac{1}{h} \left[ 2i\omega \left(  \frac{\partial_z\Psi}{\Psi} - \frac{1}{z} \right) + (ik +\partial_x)^2 \right. & &  \nonumber\\ \left. + 2ik \frac{\partial_x\Psi}{\Psi} \right] f_1
+ \frac{2A_t}{h^2} (a_t + i\omega f_2) &=& 0~, \nonumber\\
\partial_z^2 f_2 +\left[ \frac{2i\omega +h'}{h} + \frac{2\partial_z\Psi}{\Psi} - \frac{2}{z} \right] \partial_z f_2 + \frac{2}{h} \frac{\partial_x\Psi}{\Psi} \partial_x f_2+ \frac{1}{h} \left[ 2i\omega \left(  \frac{\partial_z\Psi}{\Psi} - \frac{1}{z} \right) + (ik +\partial_x)^2  \right. & &  \nonumber\\ \left.+ 2ik \frac{\partial_x\Psi}{\Psi} \right] f_2
- \frac{2 i \omega}{h} A_t f_1 - \frac{i \omega}{h} a_t -  \left(i k + \partial_x\right) a_x  - \frac{2 \partial_x \Psi}{\Psi} a_x &=& 0~, \nonumber\\
h \partial_z^2 a_t +2i\omega \partial_z a_t- \left[ \frac{i\omega h' +\omega^2}{h} - \left(i k + \partial_x\right)^2 + \frac{\Psi^2}{z^2} \right] a_t & & \nonumber\\
- \frac{\Psi^2}{z^2} \left[ 2 A_t f_1 + i \omega f_2 \right] + i \omega \left(i k + \partial_x \right) a_x &=& 0, \nonumber \\
- \frac{\Psi^2}{z^2} \partial_z f_2 + \left(i k + \partial_x\right) \partial_z a_x + \frac{i \omega}{h} \left[ \partial_z a_t
- \frac{\Psi^2}{z^2} f_2 + (ik+\partial_x) a_x + \frac{i\omega}{h} a_t \right] &=& 0, \nonumber \\
h \partial_z^2 a_x + (h' + 2i\omega ) \partial_z a_x -\frac{\Psi^2}{z^2} a_x+ \frac{\Psi^2}{z^2} \left(i k + \partial_x\right) f_2 - \frac{i \omega}{h} \left(i k + \partial_x \right) a_t &=& 0. \nonumber\\
\eea
Recall that $\Psi$ satisfies
\be z^2 \partial_z\left(\frac{h}{z^2} \, \partial_z \Psi\right) + \partial_x^2 \Psi  + \left(\frac{A_t^2}{h} + \frac{\Delta(3-\Delta)}{z^2}\right) \Psi.
= 0
\ee
The system (\ref{eq60}) has a solution which is pure gauge, given by
\be
a_x = i k e^{-i\omega z_\ast}~, \qquad a_t = - i \omega e^{-i\omega z_\ast}~, \qquad f_1 = 0~, \qquad f_2 = e^{-i\omega z_\ast}~.
\ee
Demanding regularity at the horizon as well as $a_t =0$, we obtain the boundary conditions at $z=1$,
\bea\label{eq60bc}
\left[ 2i\omega -3 \right] \partial_z f_1 + 2 \frac{\partial_x\Psi}{\Psi} \partial_x f_1+ \left[ 2i\omega \left(  \frac{\partial_z\Psi}{\Psi} - 1 \right) + (ik +\partial_x)^2 + 2ik \frac{\partial_x\Psi}{\Psi} \right] f_1
- \frac{2i\omega}{3} A_t' f_2 &=& 0~, \nonumber\\
\left[ 2i\omega -3 \right] \partial_z f_2 + 2 \frac{\partial_x\Psi}{\Psi} \partial_x f_2+ \left[ 2i\omega \left(  \frac{\partial_z\Psi}{\Psi} - 1 \right) + (ik +\partial_x)^2 + 2ik \frac{\partial_x\Psi}{\Psi} \right] f_2
&=& 0~, \nonumber\\
- \Psi^2 f_2 + (ik+\partial_x) a_x + \left[ 1- \frac{i\omega}{3} \right] \partial_z a_t &=& 0, \nonumber \\
\left[ 2i\omega -3\right] \partial_z a_x -\left[ \Psi^2 - (ik+\partial_x)^2 \right] a_x + \left(i k + \partial_x \right) \partial_z a_t &=& 0. \nonumber\\
\eea
Let us calculate the optical conductivity at low temperature in the high frequency limit. At first order, we note that the contribution of $A_t$ is negligible at low temperatures, and therefore we have $f_1 \approx 0$. Setting $k=0$ and expanding around $z=0$, we obtain
\bea\label{eq62}
\left[ \partial_z^2 + \partial_x^2 \right] f_2 +\left[ 2i\omega + \frac{2(\Delta -1)}{z} \right] \partial_z f_2 + \frac{2\partial_x\Psi}{\Psi} \partial_x f_2+ \frac{2 (\Delta -1)i\omega}{z} f_2&& \nonumber\\
- i \omega a_t -  \partial_x a_x  - \frac{2 \partial_x \Psi}{\Psi} a_x &=& 0~, \nonumber\\
\left[ \partial_z^2 + \partial_x^2 \right] a_t +2i\omega \partial_z a_t- \left[ \omega^2 + \frac{\Psi^2}{z^2} \right] a_t
- i\omega\left[ \frac{\Psi^2}{z^2} f_2 + \partial_x a_x \right] &=& 0, \nonumber \\
- \frac{\Psi^2}{z^2} \partial_z f_2 + \partial_z \partial_x a_x + i \omega \left[ \partial_z a_t
- \frac{\Psi^2}{z^2} f_2 + \partial_x a_x + i\omega a_t \right] &=& 0, \nonumber \\
\partial_z^2 a_x + 2i\omega \partial_z a_x -\frac{\Psi^2}{z^2} a_x+ \frac{\Psi^2}{z^2} \partial_x f_2 - i \omega \partial_x a_t &=& 0.
\eea
As before, this expansion is valid in the low temperature regime in which $\langle \mathcal{O}_\Delta \rangle \gtrsim 1$ and it is obtained by rescaling $z\to z / \langle \mathcal{O}_\Delta \rangle^{1/\Delta}$, and formally expanding in $1/\langle \mathcal{O}_\Delta \rangle^{1/\Delta}$. The effects of the horizon at lowest order are captured by the (exact) boundary conditions at the horizon \eqref{eq60bc}. Higher-order corrections can be systematically included, resulting in an expansion in the small parameter $1/\langle \mathcal{O}_\Delta \rangle^{1/\Delta} \sim \mathcal{O}(T_c/T)$.

 The third equation in \eqref{eq62} can be integrated to give
\be\label{eq63} i\omega a_t + \partial_x a_x = \frac{\Psi^2}{z^2} f_2 - e^{-i\omega z} \int e^{i\omega z} \partial_z \left( \frac{\Psi^2}{z^2} \right) f_2, \ee
where the constant of integration will be determined from the boundary conditions at the horizon \eqref{eq60bc}.

The remaining three equations for up to first-order corrections read
\bea\label{eq62a}
\left[ \partial_z^2 + \partial_x^2 \right] f_2 +\left[ 2i\omega + \frac{2(\Delta -1)}{z} \right] \partial_z f_2 + \frac{2\partial_x\Psi}{\Psi} \partial_x f_2+ \left[ \frac{2 (\Delta -1)i\omega}{z} - \frac{\Psi^2}{z^2} \right] f_2&& \nonumber\\
+e^{-i\omega z} \int e^{i\omega z} \partial_z \left( \frac{\Psi^2}{z^2} \right) f_2  - \frac{2 \partial_x \Psi}{\Psi} a_x &=& 0~, \nonumber\\
\left[ \partial_z^2 + \partial_x^2 \right] a_t +2i\omega \partial_z a_t- \frac{\Psi^2}{z^2} a_t
+ i\omega e^{-i\omega z} \int e^{i\omega z} \partial_z \left( \frac{\Psi^2}{z^2} \right) f_2 &=& 0, \nonumber \\
\left[ \partial_z^2 + \partial_x^2 \right] a_x + 2i\omega \partial_z a_x -\frac{\Psi^2}{z^2} a_x - \partial_x \left( \frac{\Psi^2}{z^2} \right) f_2+ e^{-i\omega z} \partial_x \int e^{i\omega z} \partial_z \left( \frac{\Psi^2}{z^2} \right) f_2 &=& 0.
\eea
Only two are needed, and the third equation can be shown to be compatible with the other two. We shall solve the first and third equations for $f_2$ and $a_x$.
Expanding in Fourier modes,
\bea
a_x &=& a_x^{(0)} + a_x^{(1)} e^{i Q x} + a_x^{(-1)} e^{- i Q x} + \dots~, \nonumber \\
f_2 &=& f_2^{(0)} + f_2^{(1)} e^{i Q x} + f_2^{(-1)} e^{- i Q x} + \dots~.
\eea
we easily see that $f_2^{(0)} =0$, and $a_x^{(0)}$ obeys the same equation as in the transverse case,
\be {a_x^{(0)}}'' + 2i\omega {a_x^{(0)}}' - {\langle\mathcal{O}_\Delta \rangle^{(0)}}^2 z^{2(\Delta -1)} a_x^{(0)} = 0. \ee
The solution can be written in the high frequency limit as (\emph{cf.}\ Eq.\ \eqref{eq23})
\be\label{eq23x} {a}_x^{(0)} = 1 + \frac{i{\langle \mathcal{O}_\Delta \rangle^{(0)}}^2}{2\omega (2\Delta -1)} z^{2\Delta -1} + \frac{{\langle \mathcal{O}_\Delta \rangle^{(0)}}^2}{(2i\omega)^{2\Delta}} \left[ \Gamma (2\Delta -1) - e^{-2i\omega z} \Gamma (2\Delta -1,- 2i\omega z) \right] + \dots \ee
For the $n= 1$ modes, we obtain
\bea\label{eq62b}
{f_2^{(1)}}'' +\left[ 2i\omega + \frac{2(\Delta -1)}{z} \right] {f_2^{(1)}}' + \left[ \frac{2(\Delta -1) i\omega}{z} -Q^2 - {\langle \mathcal{O}_\Delta \rangle^{(0)}}^2 z^{2(\Delta -1)} \right] f_2^{(1)}& & \nonumber\\
+2(\Delta -1){\langle \mathcal{O}_\Delta \rangle^{(0)}}^2 e^{-i\omega z} \int z^{2\Delta -3} e^{i\omega z} f_2^{(1)}  &=& iQ \frac{\langle\mathcal{O}_\Delta \rangle^{(1)}}{\langle\mathcal{O}_\Delta \rangle^{(0)}} a_x^{(0)}~, \nonumber\\
{a_x^{(1)}}'' + 2i\omega {a_x^{(1)}}' - \left[ Q^2 + {\langle \mathcal{O}_\Delta \rangle^{(0)}}^2 z^{2(\Delta -1)} \right] a_x^{(1)}
- \langle\mathcal{O}_\Delta \rangle^{(0)} \langle \mathcal{O}_\Delta \rangle^{(1)} z^{2(\Delta -1)} a_x^{(0)}& & \nonumber\\
+ 2i(\Delta -1)Q {\langle\mathcal{O}_\Delta \rangle^{(0)}}^2 e^{-i\omega z} \int z^{2\Delta -3} e^{i\omega z} f_2^{(1)} &=& 0~.
\eea
We note that for $\Delta =1$, the second equation has the same solution for the mode $a_x^{(1)}$ as in the transverse case, so there is no anisotropy at first order. It is also important to note that 
\be
\Re  \left[\sigma_x(\omega = 0, k=0, x)\right]  = \pi n_s \delta(\omega)~,
\ee
with $n_s \propto \langle {\cal O}_{\Delta} \rangle^{1/\Delta}$, such as given in Eq. \ref{condy}. Therefore, below $T_c$, where the scalar order parameter is non-vanishing, we obtain an infinite DC conductivity. Since translational invariance is broken in the $x$-direction, this result implies that we observe superconductivity.

To find the leading contribution to the anisotropy for $\Delta \ne 1$, approximate $a_x^{(0)} \approx 1$ in the first equation and recall that the terms involving $\Psi^2$ are subleading at high frequencies. The equation for $f_2^{(1)}$ simplifies to
\be {f_2^{(1)}}'' + \left[ 2i\omega + \frac{2(\Delta -1)}{z} \right] {f_2^{(1)}}' + \left[ \frac{2(\Delta -1)i\omega}{z} -Q^2 \right] f_2^{(1)} = iQ \frac{\langle\mathcal{O}_\Delta \rangle^{(1)}}{\langle\mathcal{O}_\Delta \rangle^{(0)}}. \ee
Its solution is given in terms of Bessel functions,
\bea f_2^{(1)} (z) &=& -\frac{iQ\pi}{2\sin\pi\nu} \frac{\langle\mathcal{O}_\Delta \rangle^{(1)}}{\langle\mathcal{O}_\Delta \rangle^{(0)}}
e^{-i\omega z} z^\nu J_{-\nu} (z\sqrt{\omega^2 -Q^2}) \left[ \mathcal{C} + \int_0^z dw w^{1-\nu} e^{i\omega w} J_{\nu} (w\sqrt{\omega^2 -Q^2}) \right] \nonumber\\
& & + \frac{iQ\pi}{2\sin\pi\nu} \frac{\langle\mathcal{O}_\Delta \rangle^{(1)}}{\langle\mathcal{O}_\Delta \rangle^{(0)}}
e^{-i\omega z} z^\nu J_{\nu} (z\sqrt{\omega^2 -Q^2}) \int_0^z dw  w^{1-\nu} e^{i\omega w} J_{-\nu} (w\sqrt{\omega^2 -Q^2}),
\label{eq70}\eea
where $\nu = \frac{3}{2} -\Delta$, and we fixed
one of the integration constants by demanding the absence of a term $z^{2\nu} = z^{3-2\Delta}$ (corresponding to $\Psi \sim z^{3-\Delta}$) near the boundary.
The remaining constant $\mathcal{C}$ is fixed by imposing the boundary condition at the horizon,
\be\label{eq71} 2i\omega {f_2^{(1)}}'(1) + \left[ 2(\Delta -1)i\omega - Q^2 \right] f_2^{(1)} (1) = 0. \ee
The other field, $a_x^{(1)}$, satisfies the equation
\be
{a_x^{(1)}}''+ 2i\omega {a_x^{(1)}}'  -Q^2 a_x^{(1)} = \mathcal{F} \ \ , \ \ \ \ \mathcal{F} = - iQ(\Delta -1){\langle \mathcal{O}_\Delta\rangle^{(0)}}^2 e^{-i\omega z} \int e^{i\omega z} z^{-2\nu} f_2^{(1)}~.
\ee
The solution is
\be a_x^{(1)} (z) = \frac{i}{2\sqrt{\omega^2 - Q^2}} e^{-i\omega z} \hat{a}_x^{(1)} (z), \ee
where
\bea \hat{a}_x^{(1)} (z) &=& \mathcal{C}' \sin \sqrt{\omega^2 -Q^2} z
+ e^{-i\sqrt{\omega^2 -Q^2} z} \int_0^z dw e^{i [ \omega + \sqrt{\omega^2 -Q^2}] w} \mathcal{F} (w) \nonumber\\
&& -  e^{i\sqrt{\omega^2 -Q^2} z} \int_0^z dw e^{i [ \omega - \sqrt{\omega^2 -Q^2}] w} \mathcal{F} (w).
\eea
One of the integration constants was fixed by demanding $a_x^{(1)} (0) =0$. Two integrations constants remain to be fixed, $\mathcal{C}'$ and the one hidden in the function $\mathcal{F}$. They are fixed by the boundary conditions at the horizon.

To expose the integration constant in the function $\mathcal{F}$, we write
\be  \mathcal{F} =  iQ e^{-i\omega z} \left[ \mathcal{C}'' + (\Delta -1){\langle \mathcal{O}_\Delta\rangle^{(0)}}^2\int_z^1 dw e^{i\omega w} w^{-2\nu} f_2^{(1)} (w) \right],
\ee
so that from Eq.\ \eqref{eq63} we can deduce
\be \mathcal{C}'' = e^{i\omega} \left[ iQa_x^{(1)} (1) - \frac{{\langle \mathcal{O}_\Delta\rangle^{(0)}}^2}{2} f_2^{(1)} (1) \right], \ee
thus expressing $\mathcal{C}''$ in terms of $\mathcal{C}'$. The latter is fixed from the last two boundary conditions in \eqref{eq60bc}, after eliminating $\partial_z a_t$ among them.

The general analytic expressions are too cumbersome to write but one may easily deduce the conductivity perpendicular to the direction of the stripes and find the anisotropy, which is non-vanishing for $\Delta \ne 1$.

\section{Near the Critical Temperature}
\label{sec:4}

As we have explained previously, the analytical techniques developed in the previous section are valid when the order parameter is a lot larger than the temperature. In order to study the system near the critical temperature, where the scalar order parameter is small, we need to be more careful in handling the boundary condition at the horizon. We can make progress in our analytical study by considering the long wavelength, small frequency regime $k, \omega \ll 1$. 

\subsection{Conductivity parallel to the direction of the stripes}

For $\sigma_y$, for convenience, let us factorize the appropriate boundary behavior
\be
A_y  = 
a_y(z,x) \, (1-z)^{-i \omega/3} \, e^{i (kx - \omega t)},
\ee
so that $a_y$ is regular at the horizon. Furthermore, we express the perturbation as a Taylor series in $\omega \sim k^2 \sim \xi^2 \ll 1$
\be
a_y = a_{y0} + a_{y\omega}\,  \omega + a_{yk} \, k^2 + a_{y \xi}\,  \xi^2 + \dots
\ee
where $\xi \sim \langle\mathcal{O}_{\Delta}\rangle^{(0)}$. Since we are expanding in $\omega$, the unwanted behavior near the horizon, $a_y \sim (1-z)^{2i\omega /3}$ will contribute singular terms at each order (\textit{e.g.}, at $\mathcal{O} (\omega)$, we obtain $a_{y\omega} \sim \frac{2i}{3} \ln (1-z)$). We exclude this behavior by demanding that our solutions admit a regular Taylor expansion in $z$ around the horizon ($z=1$).

Moreover, each field we divide into the $n=0$ mode and the $n=1, 2$ modes that are suppressed at large $Q$ (higher modes are even more suppressed). As before, we shall use a superscript to denote this hierarchy,
\be a_y (z,x) = a_y^0 (z) + a_y^{1} (z) e^{iQx} + a_y^{-1} (z) e^{-iQx} + \dots \ee
At leading order in $Q$, we have
\be
(1-z^3) \, \partial_z^2 a_{y0}^0 + \frac{h'}{h} \partial_z a_{y0}^0 = 0,
\ee
whose only solution that is regular at the horizon is the constant
\be
a_{y0}^0 (z) = a_{y0}^0 (z=0) = a_{y}^0 (z=0).
\ee
At ${\cal O}(\omega)$, we then have
\be
a^0_{y\omega}(z) = \frac{i}{3}  \, a_{y}^0 (z=0) \int_0^z \frac{dz'}{h(z')} \int_{z'}^1 \frac{1+z''- 2 (z'')^2}{1-z''} dz'',
\ee
while at ${\cal O}(k^2)$, we have
\be
a^0_{yk}(z) = - a^0_{y}(z=0) \int_0^z \frac{dz'}{h(z')}  \int_{z'}^1 dz''.
\ee
Finally, at ${\cal O}(\xi^2)$, we have
\be
a_{y\xi}^{0}  (z) = - a^0_{y}(z=0) \int_0^z \frac{dz'}{h(z')} \int_{z'}^1 dz'' {z''}^{2\Delta -2} \left[F^{(0)}(z'')\right]^2.
\ee
At sub-leading order in $Q$, all of the fields vanish except
\bea
a_{y\xi}^{\pm 1} (z) &=&  - \int_0^z \frac{dz'}{h(z')} \int_{z'}^1 dz'' {z''}^{2\Delta -2}  F^{(0)} F^{(1)}  a_y^{\pm 1} (z=0)~, \nonumber \\
a_{y\xi}^{\pm 2} (z) &=& -  \int_0^z \frac{dz'}{h(z')} \int_{z'}^1 dz'' {z''}^{2\Delta -2}  F^{(0)} F^{(2)}  a_y^{\pm 2}(z=0)~.
\eea
The dynamical conductivity parallel to the stripes is then given by
\be\label{eqnsy}
\sigma_y(\omega, k, x) = 1 + \frac{i k^2}{\omega}  +\frac{ i  \left[\langle {\cal O}_{\Delta} \rangle^{(0)} \right]^2}{\omega} \int_0^1 dz \frac{F^{(0)}}{z^{2-2\Delta}} \left[ F^{(0)} + 2 F^{(1)} \cos Qx + 2F^{(2)} \cos 2Qx \right] ~.
\ee
Once again, we obtain an infinite DC conductivity. However, as remarked earlier, this does not imply superconductivity, because of translational invariance in the $y$-direction.

\subsection{Conductivity perpendicular to the direction of the stripes}

To obtain the conductivity perpendicular to the stripes, $\sigma_x$, the perturbations that need to be turned on are
\bea
A_x  &=& \frac{a_x(z; \omega, k, x)}{k} \, (1-z)^{-i \omega/3} \,e^{i (kx - \omega t)}, \nonumber \\
A_t  &\rightarrow& A_t + 
a_t(z; \omega, k, x) \, (1-z)^{-i \omega/3} \, e^{i (kx - \omega t)}, \nonumber \\
\Psi &\rightarrow& \Psi + 
\Psi^{(0)}(z) \, f_1(z; \omega, k, x) \, (1-z)^{-i \omega/3} \, e^{i (kx - \omega t)}, \nonumber \\
\chi &=&  
\Psi^{(0)}(z) \, f_2(z; \omega, k, x) \, (1-z)^{-i \omega/3} \, e^{i (kx - \omega t)}.
\eea
Again, we have conveniently taken out a factor which captures the behavior at the horizon, so that all functions should be regular at the horizon.

The equations of motion for the $n=0$ modes are
\bea
h \, \partial_z^2 f_1^{(0)} + \left[ \frac{2 i \omega}{3} (1+z+z^2) - \frac{2+z^3}{z} + 2 h \frac{\partial_z\Psi^{(0)}}{\Psi^{(0)}} \right] \partial_zf_1^{(0)}  \qquad \qquad \qquad \qquad \qquad \qquad \nonumber \\
+ \left[ \frac{i \omega (2+z)}{3 z} + \frac{2 i \omega}{3} (1+z+z^2)\frac{\partial_z\Psi^{(0)}}{\Psi^{(0)}} - \frac{\omega^2(z+2)(4+z+z^2)}{9(1+z+z^2)} - k^2\right]f_1^{(0)} \nonumber \\
= -\frac{2}{h} A_t \left(a_t^{(0)} + i \omega f_2^{(0)} \right),\qquad
\label{01}\eea
\bea
h \, \partial_z^2 f_2^{(0)} + \left[ \frac{2 i \omega}{3} (1+z+z^2) - \frac{2+z^3}{z} + 2 h \frac{\partial_z\Psi^{(0)}}{\Psi^{(0)}} \right] \partial_zf_2^{(0)} \qquad \qquad \qquad \qquad \qquad \qquad  \nonumber \\
+ \left[ \frac{i \omega (2+z)}{3 z} + \frac{2 i \omega}{3} (1+z+z^2)\frac{\partial_z\Psi^{(0)}}{\Psi^{(0)}} - \frac{\omega^2(z+2)(4+z+z^2)}{9(1+z+z^2)} - k^2\right]f_2^{(0)} \nonumber \\
= \frac{2 i \omega}{h} A_t f_1^{(0)} + i \frac{\omega}{h} a_t^{(0)} + i a_x^{(0)}, \qquad
\label{02}\eea
\bea
h \,\partial_z^2a_t^{(0)} + \frac{2 i \omega}{3}(1+z+z^2) \partial_za_t^{(0)} - \left[k^2 + \frac{2\left(\Psi^{(0)}\right)^2}{z^2} - \frac{\omega(3i - \omega)(1+z+z^2)}{9 (1-z)}\right] a_t^{(0)}  \nonumber \\
= \frac{2\left(\Psi^{(0)}\right)^2}{z^2} \left[2 A_t f_1^{(0)} + i \omega  f_2^{(0)} \right] + \omega a_x^{(0)}, \qquad
\label{03}\eea
\be
i  \,\partial_za_x^{(0)} - \frac{\omega}{3(1-z)} a_x^{(0)} + \frac{i \omega}{h} \, \partial_za_t^{(0)} - \frac{\omega^2}{3(1-z)h} a_t^{(0)} = \frac{2\left(\Psi^{(0)}\right)^2}{z^2} \left[\partial_zf_2^{(0)}+ \frac{i \omega}{3(1-z)} f_2^{(0)} \right],
\label{04}\ee
\bea
h \, \partial_z^2a_x^{(0)} + \left[\frac{2 i \omega}{3}(1+z+z^2) -3 z^2\right] \, \partial_za_x^{(0)} \quad\quad\quad\quad\quad\quad\quad\quad\quad\quad\quad\quad\quad\quad\quad\quad\quad\quad \nonumber\\
+ \left[\frac{i \omega}{3} (1+2z) + \omega^2 \frac{8+6z+3z^2+z^3}{9(1+z+z^2)} - \frac{2\left(\Psi^{(0)}\right)^2}{z^2} \right]  a_x^{(0)}  \quad\quad\quad\quad\quad\quad\quad\quad\quad\nonumber \\
= - \frac{\omega k^2}{h} a_t^{(0)} - 2 \frac{i k^2}{z^2} \left(\Psi^{(0)}\right)^2 f_2^{(0)}~. \qquad
\label{05}
\eea
Here, we have neglected the $n=1$ mode of $A_t$, because it is exponentially suppressed at large $Q$. Thus,
\be
A_t \approx \mu (1- \delta) A_t^{(0)} \approx  \mu (1- \delta) (1-z) \equiv \mu_{\delta} (1-z)~.
\ee
Since $\omega$, $k$ and $\Psi$ are small, we can express the modes as Taylor series in $\omega$, $k^2$ and $\xi^2$ as before,
\bea
f_1^{(0)} &=& f_{10}^{(0)} + f_{1\omega}^{(0)} \omega + f_{1k}^{(0)} k^2+ f_{1\xi}^{(0)} \xi^2+ \dots \nonumber \\
f_2^{(1)} &=& f_{20}^{(0)} +  f_{2\omega}^{(0)} \omega + f_{2k}^{(0)} k^2+ f_{2\xi}^{(0)} \xi^2+ \dots\nonumber \\
a_t^{(1)} &=& a_{t0}^{(0)} +  a_{t\omega}^{(0)} \omega + a_{tk}^{(0)} k^2+ a_{t\xi}^{(0)} \xi^2+ \dots \nonumber \\
a_x^{(1)} &=& a_{x0}^{(0)} +  a_{x\omega}^{(0)}\omega + a_{xk}^{(0)} k^2+ a_{x\xi}^{(0)} \xi^2+ \dots
\eea
We shall solve the field equations starting from the horizon, so that all integration constants will be of zeroth order there. The boundary values will then be determined in terms of these horizon parameters.

At zeroth order, we obtain from Eqs. \eqref{03}, \eqref{04} and \eqref{05},
\be
\partial_z^2 a_{t0}^{(0)} = \partial_za^{(0)}_{x0} = 0,
\ee
which is solved by
\be
a_{t0}^{(0)} (z) = \alpha_0 (1-z), \ \ , \ \ \ \
a_{x0}^{(0)}  (z) = \tilde{\alpha}_0~.
\ee
The two parameters are related by demanding regularity of $a_{x\omega}^{(0)}$ in Eq. \eqref{04}
\be
\tilde{\alpha}_0 = -i\alpha_0,
\ee
and thus, we obtain
\be
a_{x\omega}^{(0)}  (z) = - \frac{\alpha_0}{3} \int_z^1 dz'  \frac{2+z}{1+z+z^2}.
\ee

From Eq. \eqref{05}, we also obtain
\be
a_{xk}^{(0)}  (z) = 0, \ \ , \ \ \ \
a_{x\xi}^{(0)}  (z) = i\alpha_0 \int_0^z \frac{dz'}{h(z')} \int_{z'}^1 dz'' {z''}^{2\Delta -2} \left[F^{(0)}(z'')\right]^2.
\ee
From Eq. \eqref{04} at $\mathcal{O} (\xi^2)$, or Eq. \eqref{02} at zeroth order, we then deduce
\be
f_{20}^{(0)} (z) = \beta_{20} -\alpha_0 \int_0^z dz' \frac{{z'}^{2-2\Delta}}{h(z')\left[F^{(0)}(z')\right]^2} \int_{z'}^1 dz''\, \frac{\left[F^{(0)}(z'')\right]^2}{ {z''}^{2-2\Delta}},
\ee
where the integration constant $\beta_{20}$ will be determined later.

Next, let us take another look at Eq. \eqref{03} to obtain
\be
a_{t\omega}^{(0)} (z) = - \frac{i\alpha_0}{3} \int_z^1 dz' \int_{z'}^1 dz'' \frac{2+z''}{1+z''+{z''}^2} \ \ , \ \ \ \
a_{tk}^{(0)} (z) = \alpha_0 \int_z^1 dz' \int_{z'}^1 \frac{dz''}{1+z''+{z''}^2}.
\ee
Now, we can solve Eq. \eqref{01} at zeroth order,
\be
f_{10}^{(0)}(z) = \beta_{10} + 2\alpha_0 \int_0^z dz' \frac{{z'}^{2-2\Delta}}{h(z')\left[F^{(0)}(z')\right]^2} \int_{z'}^1 dz''\frac{A_t(z'')}{1+z''+{z''}^2} \frac{\left[F^{(0)}(z'')\right]^2}{ {z''}^{2-2\Delta}},
\ee
where the integration constant $\beta_{10}$ will also be determined shortly. We can then obtain
\be
a_{t\xi}^{(0)} (z) = \int_z^1 dz' \int_{z'}^1 dz'' \frac{\left[F^{(0)}(z'')\right]^2}{{z''}^{2-2\Delta}} \frac{\alpha_0 + 2 \mu_{\delta} f_{10}^{(0)}(z'')}{1+z''+{z''}^2} 
\ee
from Eq. \eqref{03}.

At ${\cal O}(\omega)$, Eqs. \eqref{01} and \eqref{02} yield
\be
f_{i\omega}^{(0)}(z) = \beta_{1\omega} + \int_0^z dz'  \frac{ {z'}^{2-2\Delta}}{h(z')\left[F^{(0)}(z')\right]^2} \int_{z'}^1 \frac{dz''}{h(z'')} \frac{\mathcal{F}_i(z'')}{ {z''}^{2-2\Delta} }\ ,
\ee
for $i=1, 2$, where
\bea
 \mathcal{F}_1 (z) &=& 2A_t \left( a_{t\omega}^{(0)} + if_{20}^{(0)}\right)\nonumber\\
& & + \frac{i}{3} h \left[ 2(1+z+z^2) \partial_zf_{10}^{(0)} + \left( - \frac{2+z}{z} + 2(1+z+z^2) \frac{\partial_z\Psi^{(0)}}{\Psi^{(0)}} \right) f_{10}^{(0)} \right],
\nonumber\\
\mathcal{F}_2 (z) &=& -i\left(a_{t0}^{(0)} + ha^{(0)}_{x\omega}\right) -2i A_t f^{(0)}_{10} \nonumber\\
& & + \frac{i}{3} h \left[ 2(1+z+z^2) \partial_zf^{(0)}_{20} + \left( - \frac{2+z}{z} + 2(1+z+z^2) \frac{\partial_z\Psi^{(0)}}{\Psi} \right) f^{(0)}_{20} \right].
\eea
 We are now ready to determine the integration constants. This is done by forbidding the bulk scalar field to generate source terms for the scalar order parameter of the boundary field theory. In other words, we are going to determine the integration constants by excluding the unwanted behavior $f_{1,2} \sim z^{3-2\Delta}$ at the boundary $(z=0)$. Here, we need to distinguish between two cases, $\Delta \ne 1$ and $\Delta =1$.

For $\Delta \ne 1$, to isolate the unwanted behavior, we write
\be 
f^{(0)}_{10} = \hat{f}_{10} + \check{f}_{10},
\ee
where
\bea
\hat{f}_{10} &=& \beta_{10} - 2\alpha_0 \int_0^z dz' \frac{{z'}^{2-2\Delta}}{h(z')\left[F^{(0)}(z')\right]^2} \int_0^{z'} dz''\frac{A_t(z'')}{1+z''+{z''}^2} \frac{\left[F^{(0)}(z'')\right]^2}{ {z''}^{2-2\Delta}}, \nonumber \\
 \check{f}_{10} &=& 2\alpha_0 \int_0^z dz' \frac{{z'}^{2-2\Delta}}{h(z')\left[F^{(0)}(z')\right]^2} \int_0^{1} dz''\frac{A_t(z'')}{1+z''+{z''}^2} \frac{\left[F^{(0)}(z'')\right]^2}{ {z''}^{2-2\Delta}}.
\eea
Hereupon, we use the notation $\hat{}$ and $\check{}$ to describe the wanted and unwanted near boundary behavior, respectively, so that $\hat{g} \sim z^0$ while $\check{g} \sim z^{3-2\Delta}$.

Similarly, for $f_{20}$, we have
\be 
f^{(0)}_{20} = \hat{f}_{20} + \check{f}_{20},
\ee
where
\bea
\hat{f}_{20} &=& \beta_{20} + \alpha_0 \int_0^z dz' \frac{{z'}^{2-2\Delta}}{h(z')\left[F^{(0)}(z')\right]^2} \int_0^{z'}dz''\, \frac{\left[F^{(0)}(z'')\right]^2}{ {z''}^{2-2\Delta}}, \nonumber \\
 \check{f}_{20}&=& - \alpha_0 \int_0^z dz' \frac{{z'}^{2-2\Delta}}{h(z')\left[F^{(0)}(z')\right]^2} \int_0^{1}dz''\, \frac{\left[F^{(0)}(z'')\right]^2}{ {z''}^{2-2\Delta}}.
\eea
Next, we split
\be
\mathcal{F}_i = \hat{\mathcal{F}}_i + \check{\mathcal{F}}_i \,,
\ee
for $i = 1, 2,$ with
\bea
\hat{\mathcal{F}}_1  &=& 2A_t \left( a_{t\omega}^{(0)} + i\hat{f}_{20}^{(0)}\right)\nonumber\\
& & + \frac{i}{3} h \left[ 2(1+z+z^2) \partial_z\hat{f}_{10}^{(0)} + \left( - \frac{2+z}{z} + 2(1+z+z^2) \frac{\partial_z\Psi^{(0)}}{\Psi^{(0)}} \right) \hat{f}_{10}^{(0)} \right], \nonumber\\
\check{ \mathcal{F}}_1 &=& 2i A_t  \check{f}_{20}^{(0)} + \frac{i}{3} h \left[ 2(1+z+z^2) \partial_z\check{f}_{10}^{(0)} + \left( - \frac{2+z}{z} + 2(1+z+z^2) \frac{\partial_z\Psi^{(0)}}{\Psi^{(0)}} \right) \check{f}_{10}^{(0)} \right],\qquad
\eea
and
\bea
\hat{\mathcal{F}}_2  &=&-i\left(a_{t0}^{(0)} + ha^{(0)}_{x\omega} \right) -2i A_t \hat{f}^{(0)}_{10} \nonumber\\
& & + \frac{i}{3} h \left[ 2(1+z+z^2) \partial_z\hat{f}^{(0)}_{20} + \left( - \frac{2+z}{z} + 2(1+z+z^2) \frac{\partial_z\Psi^{(0)}}{\Psi} \right) \hat{f}^{(0)}_{20} \right], \nonumber\\
\check{\mathcal{F}}_2  &=& -2i A_t \check{f}^{(0)}_{10} + \frac{i}{3} h \left[ 2(1+z+z^2) \partial_z\check{f}^{(0)}_{20} + \left( - \frac{2+z}{z} + 2(1+z+z^2) \frac{\partial_z\Psi^{(0)}}{\Psi} \right) \check{f}^{(0)}_{20} \right].\qquad
\eea
Therefore,
\be
\check{f}_{i\omega} (z)= \int_0^z dz' \frac{ {z'}^{2-2\Delta}}{h(z')\left[F^{(0)}(z')\right]^2} \int_{z'}^1 \frac{dz''}{h(z'')}   \frac{\check{\mathcal{F}}_i(z'')}{{z''}^{2-2\Delta}}
+ \int_0^z dz' \frac{ {z'}^{2-2\Delta}}{h(z')\left[F^{(0)}(z')\right]^2} \int_0^1 \frac{dz''}{h(z'')}  \frac{\hat{\mathcal{F}}_i(z'')}{{z''}^{2-2\Delta}}.
\ee
At lowest order, we then obtain
\be
\mathcal{A} \beta_{10}  + \mathcal{B} \beta_{20}  = \frac{2i\mathcal{C}\alpha_0}{\omega} \ \ , \ \ \ \
\mathcal{A} \beta_{20} - \mathcal{B} \beta_{10} = -\frac{i\mathcal{D}\alpha_0}{\omega},
\ee
where
\bea 
\mathcal{A} &=& \frac{1}{3} \int_0^1 dz\, z^{2\Delta -2} \left( - \frac{2+z}{z} + 2(1+z+z^2) \frac{\partial_z\Psi}{\Psi} \right),  \ \ \mathcal{B} = 2 \mu_{\delta} \int_0^1 dz \frac{z^{2\Delta -2}}{1+z+z^2}, \\
\mathcal{C} &=& \int_0^1 dz \frac{z^{2\Delta -2} A_t}{1+z+z^2} \left[F^{(0)}\right]^2 , \qquad \qquad \qquad \qquad \qquad \mathcal{D} = \int_0^1 dz \, z^{2\Delta -2} \left[F^{(0)}\right]^2.
\eea
Therefore,
\be
\beta_{i0} = \frac{i \alpha_0}{\omega} x_i,
\ee
where
\be
x_1 = \frac{2 \mathcal{A} \mathcal{C}+\mathcal{B} \mathcal{D}}{\mathcal{A}^2+ \mathcal{B}^2} \ \ , \ \ \ \
x_2 = \frac{\mathcal{A} \mathcal{D} - 2\mathcal{B} \mathcal{C}}{\mathcal{A}^2+ \mathcal{B}^2}.
\ee
For $\Delta =1$, $\beta_{i0}$ ($i=1,2$) are found by demanding the absence of a $z^2$ term in the near boundary asymptotics of $\Psi$ and $\chi$. This implies
\be 
\partial_z f_{1,2} + \frac{i\omega}{3} f_{1,2} = 0 
\ee
at $z=0$. We then have
\be
\left(\mathcal{A} + \frac{1}{3} \right)\beta_{10}  + \mathcal{B} \beta_{20}  = \frac{2i\mathcal{C}\alpha_0}{\omega} \ \ , \  \ \ \
\left(\mathcal{A} + \frac{1}{3} \right) \beta_{20} - \mathcal{B} \beta_{10} = -\frac{i\mathcal{D}\alpha_0}{\omega},
\ee
where $\mathcal{A}$, $\mathcal{B}$, $\mathcal{C}$ and $\mathcal{D}$ are evaluated at $\Delta =1$. Therefore,
\be
x_1 = \frac{2 \left(\mathcal{A}+ \tfrac{1}{3}\right) \mathcal{C}+\mathcal{B} \mathcal{D}}{\left(\mathcal{A}+ \tfrac{1}{3}\right)^2+ \mathcal{B}^2} \ \ , \ \ \ \
x_2 = \frac{\left(\mathcal{A}+ \tfrac{1}{3}\right) \mathcal{D} - 2\mathcal{B} \mathcal{C}}{\left(\mathcal{A}+ \tfrac{1}{3}\right)^2+ \mathcal{B}^2}.
\ee
At a glance, it seems that there is a discontinuity at $\Delta = 1$. However, this is not true. To see that we do in fact obtain the correct limit, let $\Delta = 1+\epsilon$. Then, we have
\bea
\mathcal{A} &=& \frac{1}{3} \int_0^1 dz\, z^{2\epsilon} \left[ - \frac{2+z}{z} + 2(1+z+z^2) \left( \frac{1+\epsilon}{z} + \frac{\partial_zF^{(0)}}{F} \right) \right] \nonumber \\
&=& \frac{1}{3} \int_0^1 dz\, z^{2\epsilon} \left[  \frac{2\epsilon+z+2z^2}{z} + 2(1+z+z^2) \frac{\partial_zF}{F} \right] + \mathcal{O} (\epsilon) \nonumber\\
&=& \frac{1}{3} \int_0^1 dz\, 2\epsilon z^{2\epsilon-1} + \mathcal{A}\Big|_{\Delta =1} + \mathcal{O} (\epsilon) \nonumber\\
&=& \frac{1}{3} + \mathcal{A}\Big|_{\Delta =1} + \mathcal{O} (\epsilon)
\eea
showing that we recover the expression used in the $\Delta =1$ case as $\epsilon\to 0$.

Next, let us consider the $n=1$ modes. For convenience, rescale $f_1^{(1)} \rightarrow f_1^{(1)}/k$, $f_2^{(1)} \rightarrow f_2^{(1)}/k$ and $a_t^{(1)} \rightarrow a_t^{(1)}/k$. Then the equations of motion are
\bea
h \partial_z^2f_1^{(1)} + \left[ \frac{2 i \omega}{3} (1+z+z^2) - \frac{2+z^3}{z} + 2 h \frac{\partial_z\Psi^{(0)}}{\Psi^{(0)}} \right] \partial_zf_1^{(1)}  \qquad \qquad \qquad \qquad \qquad \qquad \nonumber \\
+ \left[ \frac{i \omega (2+z)}{3 z} + \frac{2 i \omega}{3} (1+z+z^2)\frac{\partial_z\Psi^{(0)}}{\Psi^{(0)}} - \frac{\omega^2(z+2)(4+z+z^2)}{9(1+z+z^2)} - (k+Q)^2\right]f_1^{(1)} \nonumber \\
= -\frac{2}{h} A_t \left(a_t^{(1)} + \frac{k}{2} \frac{F^{(1)}}{F^{(0)}} a_t^{(0)} + i \omega f_2^{(1)} \right), \qquad
\label{eq-11}\eea
\bea
h \partial_z^2f_2^{(1)} + \left[ \frac{2 i \omega}{3} (1+z+z^2) - \frac{2+z^3}{z} + 2 h \frac{\partial_z\Psi^{(0)}}{\Psi^{(0)}} \right] \partial_zf_2^{(1)} \qquad \qquad \qquad \qquad \qquad \qquad  \nonumber \\
+ \left[ \frac{i \omega (2+z)}{3 z} + \frac{2 i \omega}{3} (1+z+z^2)\frac{\partial_z\Psi^{(0)}}{\Psi^{(0)}} - \frac{\omega^2(z+2)(4+z+z^2)}{9(1+z+z^2)} - (k+Q)^2\right]f_2^{(1)} \nonumber \\
= \frac{2 i \omega}{h} A_t f_1^{(1)} + i \left[\frac{\omega}{h} a_t^{(1)} + \left(k+ Q\right)a_x^{(1)}\right] + \frac{k}{2} \frac{F^{(1)}}{F^{(0)}}  \left[\frac{i \omega}{h} a_t^{(0)} + a_x^{(0)} \right] - i Q \frac{F^{(1)}}{F^{(0)}} a_x^{(0)}, \qquad
\label{eq-12}\eea
\bea
h \partial_z^2a_t^{(1)} + \frac{2 i \omega}{3}(1+z+z^2) \partial_za_t^{(1)} - \left[(k+Q)^2 + \frac{2\left(\Psi^{(0)}\right)^2}{z^2} - \frac{\omega(3i - \omega)(1+z+z^2)}{9 (1-z)}\right] a_t^{(1)}  \nonumber \\
= \frac{4 k\Psi^{(0)}\Psi^{(1)}}{z^2} \left[ a_t^{(0)} +  A_t f_1^{(0)} \right] + \frac{4\left(\Psi^{(0)}\right)^2 A_t f_1^{(1)}}{z^2}  \quad\quad\quad\quad\quad\quad \nonumber \\
+ \frac{2i \omega \Psi^{(0)}}{z^2} \left[\Psi^{(0)} f_2^{(1)} + k \Psi^{(1)} f_2^{(0)} \right] + \omega \left(k+ Q \right) a_x^{(1)}, \qquad
\label{eq-13}\eea
\bea
i \left(k+ Q\right) \partial_za_x^{(1)} - \frac{\omega k}{3(1-z)} a_x^{(1)} + \frac{i \omega}{h} \partial_za_t^{(1)} - \frac{\omega^2}{3(1-z)h} a_t^{(1)} \qquad \qquad \qquad \qquad \qquad  \qquad \nonumber \\
= \frac{2\left(\Psi^{(0)}\right)^2}{z^2} \left[\partial_zf_2^{(1)} + \frac{i \omega}{3(1-z)} f_2^{(1)} \right] \quad\quad\quad\quad\quad\quad\quad\quad\quad\quad\quad\quad\quad\quad \nonumber\\
+ \frac{2\Psi^{(0)}\Psi^{(1)}}{z^2} \left[\partial_zf_2^{(1)} + \frac{i \omega}{3(1-z)} f_2^{(1)} \right]  - \frac{2k \Psi^{(1)}}{z^2} f_2^{(0)} \partial_z\Psi^{(0)},\qquad
\label{eq-14}\eea
\bea
h \partial_z^2a_x^{(1)} + \left[\frac{2 i \omega}{3}(1+z+z^2) -3 z^2\right] \partial_za_x^{(1)} \quad\quad\quad\quad\quad\quad\quad\quad\quad\quad\quad\quad\quad\quad \nonumber\\
+ \left[\frac{i \omega}{3} (1+2z) + \omega^2 \frac{8+6z+3z^2+z^3}{9(1+z+z^2)} - \frac{2\left(\Psi^{(0)}\right)^2}{z^2} \right]  a_x^{(1)}  \quad\quad\quad\quad\quad\quad\nonumber \\
= - \frac{\omega (k+Q)}{h} a_t^{(1)} - \frac{2i (k+Q)}{z^2} \left(\Psi^{(0)}\right)^2 f_2^{(1)} \quad\quad\quad\quad \nonumber\\
+ \frac{2i k(Q-k)}{z^2} \Psi^{(0)} \Psi^{(1)} f_2^{(0)} + \frac{4 \Psi^{(0)} \Psi^{(1)}}{z^2} a_x^{(0)}. \qquad
\label{eq-15}
\eea
As with the $n=0$ modes, we expand the $n=1$ modes as
\bea
f_1^{(1)} &=& f_{10}^{(1)} +  f_{1\omega}^{(1)} \omega + f_{1k}^{(1)} k^2+ f_{1\xi}^{(1)} \xi^2 + \dots \nonumber \\
f_2^{(1)} &=& f_{20}^{(1)} +  f_{2\omega}^{(1)} \omega + f_{2k}^{(1)} k^2+ f_{2\xi}^{(1)} \xi^2 + \dots\nonumber \\
a_t^{(1)} &=& a_{t0}^{(1)} +  a_{t\omega}^{(1)} \omega + a_{tk}^{(1)} k^2+ a_{t\xi}^{(1)} \xi^2 + \dots \nonumber \\
a_x^{(1)} &=& a_{x0}^{(1)} +  a_{x\omega}^{(1)} \omega + a_{xk}^{(1)} k^2+ a_{x\xi}^{(1)} \xi^2 + \dots
\eea
in the small $\omega$, $k$ and $\xi$ regime. Then Eq. \eqref{eq-13}, along with the boundary conditions $a_{t0}^{(1)} (z = 0) = 0$ and $a_{t0}^{(1)} (z = 1) = 0$, give us $a_{t0}^{(1)} (z) = 0$. Similarly, $a_{t\omega}^{(1)} (z = 0) = a_{tk}^{(1)} (z = 0) =  0$. Eq. \eqref{eq-14}, along with the boundary condition $a_{x0}^{(1)} (z = 0) = 0$, and demanding regularity at the horizon, gives us $a_{x0}^{(1)} (z) = 0$. Similarly, $a_{x\omega}^{(1)} (z = 0) = a_{xk}^{(1)} (z = 0) = 0$.

In order to solve Eq. \eqref{eq-12} at leading order, we do a coordinate transformation $\tilde{z} = Q z$, and keep the leading terms in $Q$ to obtain
\be
\tilde{z}^{2- 2\Delta} \, \partial_{\tilde{z}} \left(\tilde{z}^{2\Delta-2}\, \partial_{\tilde{z}} f^{(1)}_{20}\right) -  f^{(1)}_{20} = -  \frac{\alpha_0}{Q} \, \frac{\langle {\cal O}_{\Delta} \rangle^{(1)}}{\langle {\cal O}_{\Delta} \rangle^{(0)}}.
\ee
The solutions to the homogeneous part of this differential equation can be expressed as linear combinations of Bessel functions
\be
\tilde{z}^{\frac{3-2\Delta}{2}} \,  J_{\frac{2\Delta-3}{2}}(-i \tilde{z}) \qquad {\rm and} \qquad \tilde{z}^{\frac{3-2\Delta}{2}} \,  Y_{\frac{2\Delta-3}{2}}(-i \tilde{z}).
\ee
However, the latter will give rise to a source term at the boundary and thus, the solution that satisfies all necessary boundary conditions in the original radial coordinate is given by
\be
f^{(1)}_{20} =  \frac{\alpha_0}{Q} \, \frac{\langle {\cal O}_{\Delta} \rangle^{(1)}}{\langle {\cal O}_{\Delta} \rangle^{(0)}} \mathcal{J}_\Delta (-iQz)~,
\ee
where
\be \mathcal{J}_\Delta (u) \equiv 1+ \Gamma\left(\Delta-\frac{1}{2}\right) \left(\frac{u}{2} \right)^{\frac{3-2 \Delta}{2}} J_{\frac{2\Delta-3}{2}}(u) ~. \ee
From Eq. \eqref{eq-15}, we then obtain
\be
a_{x \xi}^{(1)} = i \alpha_0 \int_0^z \frac{dz'}{h(z')} \int_{z'}^1 dz'' \, {z''}^{2 \Delta - 2} \left(F^{(0)} \right)^2 \left[ \frac{F^{(1)}}{F^{(0)}} +
2 \ \frac{\langle {\cal O}_{\Delta} \rangle^{(1)}}{\langle {\cal O}_{\Delta} \rangle^{(0)}}  \mathcal{J}_\Delta (-iQz) \right]~.
\ee
At zeroth order, Eq. \eqref{eq-11}, along with the boundary conditions, and demanding regularity at the horizon, yields $f_{10}^{(1)} = 0$. Substituting this into Eq. \eqref{eq-13} at ${\cal O} (\xi^2)$, we get $a_{t\xi}^{(1)}=0$. 

Moving on to the other modes, the $n=-1$ modes are similar, and the $n= \pm2$ modes  are obtained by substituting $Q \rightarrow 2 Q$, $\langle {\cal O}_{\Delta} \rangle^{(1)} \rightarrow  \langle {\cal O}_{\Delta} \rangle^{(2)}$ and $F^{(1)} \rightarrow F^{(2)}$ into the results for $n=1$ modes.

The boundary values of the perturbations of the gauge fields are then given by
\bea
a_t(0) &\approx& \alpha_0 + a^{(0)}_{t\omega} (0)  \omega+ a^{(0)}_{tk} (0) k^2 + a_{t\xi}^{(0)} (0) \xi^2 \approx \alpha_0 \left[ 1 +  \frac{2i C}{\omega} \left(\langle {\cal O}_{\Delta} \rangle^{(0)} \right)^2  \right], \nonumber \\
a_x(0) &\approx& \tilde{\alpha}_0 + a^{(0)}_{x\omega} (0) \omega + a^{(0)}_{x\xi} (0) \xi^2 \approx -i\alpha_0,
\eea
where
\be
C= \mu_{\delta} \, x_1 \int_0^1 dz \int_z^1 \frac{dz'}{h}  \frac{\left[F^{(0)}(z')\right]^2}{{z'}^{2-2\Delta}}.
\ee
Clearly, this is not a general solution, because the boundary values of $a_t$ and $a_x$ are not independent of each other. To correct this, we need to mix in an arbitrary amount of pure gauge solution
\be 
a_t (1-z)^{-i \omega/3}= -i\omega \gamma_0 \qquad {\rm and} \qquad a_x (1-z)^{-i \omega/3}= ik^2 \gamma_0.
\ee
We then have
\be
a_t(0) = \alpha_0\left[ 1 +  \frac{2i C}{\omega} \left(\langle {\cal O}_{\Delta} \rangle^{(0)} \right)^2  \right]- i\omega \gamma_0\ \ , \ \ \ \
a_x(0) = -i\alpha_0  + ik^2 \gamma_0.
\ee
Solving for the arbitrary parameters, we obtain
\be
\alpha_0 = \frac{k^2a_t(0) + \omega a_x(0)}{k^2\left[ 1 +  \frac{2i C}{\omega} \left(\langle {\cal O}_{\Delta} \rangle^{(0)} \right)^2  \right]-i\omega}\ \ , \ \ \ \
\gamma_0 = \frac{a_t(0) - i  \left[ 1 +  \frac{2i C}{\omega} \left(\langle {\cal O}_{\Delta} \rangle^{(0)} \right)^2  \right] a_x(0)}{k^2\left[ 1 +  \frac{2i C}{\omega} \left(\langle {\cal O}_{\Delta} \rangle^{(0)} \right)^2  \right]-i\omega}.
\ee
We can now read the retarded Green functions from the on-shell action. The (charge) density-density retarded correlation function is given by
\bea
\big\langle \rho \, \rho \big\rangle (\omega, k , x) &=& \frac{k^2}{k^2\left(\omega +  2i C \left[\langle {\cal O}_{\Delta} \rangle^{(0)} \right]^2  \right) - i\omega^2} \Bigg[\omega +2 i \, \mu_{\delta} \, x_1 \left[\langle {\cal O}_{\Delta} \rangle^{(0)} \right]^2 
\int_0^1 \frac{dz}{1+z+z^2} \frac{\left[F^{(0)}\right]^2}{z^{2-2\Delta}} \nonumber \\
&& \qquad \qquad + \frac{\pi}{3 \sqrt{3}}\, \omega  k^2- \frac{i (\pi + 2\sqrt{3} + \sqrt{3} \log 3)}{6 \sqrt{3}} \,  \omega^2  + \frac{2C}{3}\, \omega \left[\langle {\cal O}_{\Delta} \rangle^{(0)} \right]^2 \Bigg],
\eea
with poles at
\be
\omega = -\frac{ik^2}{2} \left[ 1 \pm \sqrt{1 - 4C \frac{\left[\langle {\cal O}_{\Delta} \rangle^{(0)} \right]^2}{k^2}} \right].
\ee
When $\langle {\cal O}_{\Delta} \rangle^{(0)} \ll k$, we get a diffusion pole
\be 
\omega = -ik^2,
\ee
showing that the behavior of the system is closer to a normal fluid then a superfluid, while when $\langle {\cal O}_{\Delta} \rangle^{(0)} \gg k$, the behavior is closer to that of a superfluid and we get poles at
\be
\omega = \pm \sqrt{C} \,  \langle {\cal O}_{\Delta} \rangle^{(0)} \, k,
\ee
which is identified as the second sound with the speed of the second sound given by
\be 
v = \sqrt{C} \,  \langle {\cal O}_{\Delta} \rangle^{(0)}.
\ee
We can also obtain the thermodynamic susceptibility,
\bea
\partial_{\mu}^2 P &=& \lim_{k \rightarrow 0} \big\langle \rho \, \rho \big\rangle (0, k , x) \nonumber \\
&=& \frac{\mu_{\delta} \, x_1}{C}  
\int_0^1 \frac{dz}{1+z+z^2} \frac{\left[F^{(0)}\right]^2}{z^{2-2\Delta}}, 
\eea
which is positive definite.

Next, we can read the current-current retarded correlation function from the on-shell action, and then take the limit $k\to 0$ to get the optical conductivity in the direction of the stripes,
\be
\sigma_x(\omega, k=0, x) = 1 +i\frac{n_{s,x}}{\omega} + \dots ~,\ee
where
\bea n_{s,x} &=&  \left[\langle {\cal O}_{\Delta} \rangle^{(0)} \right]^2 \int_0^1 dz \frac{F^{(0)}}{z^{2-2\Delta}} \left[ F^{(0)} + 2 F^{(1)} \cos Qx + 2 F^{(2)} \cos 2Qx \right] \nonumber \\
& & +\,  4\,  \langle {\cal O}_{\Delta} \rangle^{(0)}\langle {\cal O}_{\Delta} \rangle^{(1)} \int_0^1 dz \frac{\left[F^{(0)}\right]^2}{z^{2-2\Delta}} \mathcal{J}_\Delta (-iQz) \cos Qx \nonumber\\
& & +\,  4 \, \langle {\cal O}_{\Delta} \rangle^{(0)}\langle {\cal O}_{\Delta} \rangle^{(2)} \int_0^1 dz \frac{\left[F^{(0)}\right]^2}{z^{2-2\Delta}}
\mathcal{J}_\Delta (-2iQz) \cos 2Qx \  .
\eea
It is easy to check that $n_s$ is real. It is related to the DC conductivity via \eqref{eq38a}. Here, again we see that the superfluid density $n_{s,x} \ne 0$ below the critical temperature, and vanishes as the scalar order parameter vanishes at $T=T_c$. Since there is no translational invariance in the $x$-direction, this corresponds to the onset of superconductivity at the critical temperature $T_c$.

Comparing with the conductivity parallel to the stripes \eqref{eqnsy} , we see that the optical conductivity is anisotropic,
\be \sigma_x(\omega ,x) - \sigma_y(\omega,x) = i \frac{\varpi_0}{\omega} + \dots \ee
where
\bea \varpi_0 &=&  4 \, \langle {\cal O}_{\Delta} \rangle^{(0)}\langle {\cal O}_{\Delta} \rangle^{(1)} \int_0^1 dz \frac{\left[F^{(0)}\right]^2}{z^{2-2\Delta}} \mathcal{J}_\Delta (-iQz) \cos Qx \nonumber\\
& & + \, 4\,  \langle {\cal O}_{\Delta} \rangle^{(0)}\langle {\cal O}_{\Delta} \rangle^{(2)} \int_0^1 dz \frac{\left[F^{(0)}\right]^2}{z^{2-2\Delta}}
\mathcal{J}_\Delta (-2iQz) \cos 2Qx 
\eea
is non-vanishing. A more complete study of the anisotropy and comparison with experiment would be of interest.

\section{Conclusion}
\label{sec:5}

In this article, we have calculated the conductivity of a strongly coupled striped superconductor at large modulation regime, both in the directions perpendicular and parallel to the stripes. In the dual gravity picture, this is done by adding perturbations in term of vector potential on top the solution of Einstein-Maxwell-scalar found in Ref. \cite{Hutasoit:2012fk}.

We found that below the critical temperature, the imaginary part of the optical conductivity has a pole at $\omega =0$, which implies that the real part of the conductivity contains a term that is proportional to $\delta(\omega)$, thus establishing that this model indeed exhibits superconductivity below the critical temperature.

At $T \lesssim T_c$, we found that the conductivity is anisotropic and it exhibits $1/\omega$ behavior at small frequency. This result is valid for $\omega/T \sim \langle {\cal O}_{\Delta} \rangle^{2}/T^{2\Delta}$ small. At low temperature $T \ll T_c$, we were only able to explicitly write down $\sigma_x$ for $\Delta = 1$, and in this case we found that the optical conductivity is isotropic. This is valid at low enough temperature where the bulk electrostatic potential can be neglected. As the anisotropy at low temperature has been measured \cite{Lee:2005fk}, our results seem to rule out $\Delta = 1$. As we have seen that the anisotropy is non-vanishing for $\Delta \ne 1$, it will be interesting to study this model at low temperature for different values of $\Delta$ numerically and compare it to the $1/\omega^{1.6}$ behavior observed experimentally at large frequency. 

It will also be interesting to study the anisotropy at $T>T_c$ by introducing fermion into this holographic set-up and see whether there is something akin to the ``dimensional crossover" as studied in \cite{doi:10.1021/cr030647c}.

Lastly, we would like to note that the anisotropy in the optical conductivity we are studying in this article emerges from the fact that the translational symmetry is broken in one direction. It is not related to the point group symmetry of the order parameter, as can be seen from the fact that our order parameter is a scalar, \textit{i.e.}, an $s$-wave order parameter, albeit one with broken translational symmetry. The fact that the anisotropy of the conductivity is not related to the anisotropy of the order parameter (if there is any) can also be deduced from the experimental observations in cuprates where the anisotropy of the conductivity is also observed above the critical temperature \cite{Dordevic:fk}.

\acknowledgments{We thank S.\ Papanikolaou for discussions. J.\ H.\ is supported by West Virginia University start-up funds. The work of G.\ S.\ and J.\ T.\ is supported in part by the Department of Energy under Grant No.\ DE-FG05-91ER40627. }

\bibliographystyle{kp}
\bibliography{References}
\end{document}